\documentclass{aa}
\usepackage{graphics}
\newcommand{\msol}{\,{\rm M_\odot}}
\begin{document}

%
   \title{The origin of the young stellar population in the solar neighborhood
--- a link to the formation of the Local Bubble?}

   \author{T. W. Bergh\"ofer
          \inst{1}
          \and
          D. Breitschwerdt\inst{2}
          }
   \offprints{T. W. Bergh\"ofer}

   \institute{DESY Hamburg, Notkestra{\ss}e 85, D-22607 Hamburg,
        Germany\\ email: thomas.berghoefer@desy.de
        \and
        Max-Planck-Institut f\"ur Extraterrestrische Physik,
        Giessenbachstr.\ 1, D-85740 Garching, Germany\\
        email: breitsch@mpe.mpg.de}

   \date{Received September 30, 2001; accepted April 24, 2002}

   \titlerunning{Young stars and the Local Bubble}

\abstract{
We have analyzed the trajectories of moving stellar groups in the
solar neighborhood in an attempt to estimate the number of supernova
explosions in our local environment during the past 20 million years.
Using Hipparcos stellar distances and the results of kinematical
analyses by Asiain et al. (1999a) on the Pleiades moving groups, we
are able to show that subgroup B1, consisting of early type B stars
up to $10\, {\rm M}_\odot$, but lacking more massive objects,
has passed through the local interstellar medium within less than 100 pc.
Comparing the stellar content of B1 with the initial mass function
derived from the analysis of galactic OB associations, we estimate the
number of supernova explosions and find that about 20 supernovae must have 
occurred during the past $\sim 10 - 20$ million years, which is suggested  
to be the age of the Local Bubble; the age of the star cluster is about 
$\sim 20 - 30$ million years. For the first time, this provides 
strong evidence that the Local Bubble must have been created and shaped by
multi-supernova explosions and presumably been reheated more than 1 
million years ago, consistent with recent findings of an excess of 
${}^{60}{\rm Fe}$ in a deep ocean ferromanganese crust. 
Calculating similarity solutions of an expanding 
superbubble for time-dependent energy input, we show that the number of 
explosions is sufficient to explain the size of the Local Bubble. 
The present energy input rate is about 
$\dot E_{\rm SN} \sim 5 \times 10^{36}$ erg/s, in good agreement with the 
estimated local soft X-ray photon output rate.
It seems plausible that the origin of the Local
Bubble is also linked to the formation of the Gould Belt, which originated 
about 30-60 Myrs ago.
}
\maketitle


\section{Introduction}

Our solar system is embedded in a medium of low H{\sc i} column density 
(e.g.\ Frisch \& York 1983) largely filled with hot plasma 
radiating in the soft X-rays (McCammon et al. 1983). 
The obvious anticorrelation between 
neutral gas and X-ray emitting plasma on large angular scales 
has been interpreted in terms of an elongated local hot cavity 
(Tanaka \& Bleeker 1977, Sanders et al., 1977; Snowden et al., 1990) with 
an extension between 80 and 200 pc, now commonly known as the 
Local Bubble (LB). Recently, based on the spatial distribution of stellar
EUV sources detected with EUVE and subsequent Na{\sc i} absorption line 
studies (Sfeir et al. 1999), Welsh et al. (1999) suggested that the LB shows a 
``chimney like'' structure 
with no definite border towards high Galactic latitude in the northern 
direction rather than a full bubble. Despite the detailed mappings of the 
LB and our knowledge about its morphology, its origin is still a mystery. 

Although it has often been suggested that the LB is the result of one or more 
supernova (SN) explosions (Cox \& Anderson 1982, Innes \& Hartquist 1984), 
there is no {\em direct} evidence for this 
hypothesis. Unlike in the neighboring Loop I superbubble, 
with still ongoing star formation and 
about 40 SN explosions to occur in the future (Egger 1998), no 
obvious active cluster of early type stars resides inside the LB. 
Moreover, the analysis of the line of sight in the direction of the star 
$\beta$ CMa, which extends over 200~pc, seems to indicate that the cavity is 
not homogeneously filled with X-ray emitting plasma (Gry et al. 1985, 
Welsh et al. 1998) and a non-negligible part of the LB material must exist in 
a warm phase with a temperature of several 10$^4$~K.
This rules out a LB formation by one single SN explosion since an expanding 
pressure driven remnant would
always fill the cavity inside a swept-up shell.  

On the other hand, there is also support for an isotropic 
nature of the LB emission, coming from EUVE observations of cloud 
shadows in the local ISM. Based on these observations, Bergh\"ofer et al. 
(1998) found evidence that the pressure of the hot interstellar medium is 
the same in three different directions near the Galactic plane.
Therefore, it seems appropriate to distinguish between the local cavity as 
the region largely devoid of H{\sc i} gas and the present day LB as the 
X-ray emitting region.  
These ambiguous observational results have consequently led to a variety of
different ideas concerning the origin of the LB. 

Some time ago the mysterious $\gamma$-ray source Geminga seemed to be the
most promising candidate for a recent nearby SN explosion, after having
been identified with an X-ray millisecond pulsar (Gehrels \& Chen, 1993).
However, HST observations have provided a parallactic distance of
$\approx 160$ pc (Caraveo et al., 1996). Therefore, due to its high proper 
motion, Geminga most probably originated $3.4 \times 10^5$ yr ago in the Orion 
region and is thus not related to the present day EUV and soft X-ray emission
of the LB.

In another scenario, the superbubble picture of the LB is disputed, and
the LB is conceived as being related to a local interarm region
between the Sagittarius and the Perseus spiral arms of the Galaxy
(Bochkarev 1987; Frisch 1995, 1996). It is assumed that during 
different epochs of star formation in the Sco-Cen association hot gas was 
generated and subsequently swept into the surrounding inhomogeneous medium. 
While some morphological
features of the local ISM  can be explained by such a model, 
 it cannot account for the distinct X-ray shadow observed
towards the Loop I bubble in the
ROSAT PSPC (Position Sensitive Proportional Counter) data 
(Egger \& Aschenbach 1995), which is most naturally explained
by an interaction ring between Loop I and the LB. The existence of a
``wall'' between these two bubbles has also been demonstrated by
stellar absorption line measurements of stars with known Hipparcos distances.
The formation of such a wall is rather difficult to explain once hot
plasma is flowing out from Loop I directly into the LB.
For a detailed discussion we refer the reader to Breitschwerdt et al. (2000).

Recently, Smith \& Cox (2001) have performed 1-D hydrodynamic simulations
to explore whether a low energy input rate in combination with a high ambient
pressure can result in the formation of a bubble like the LB. 
They find that $2-3$ SN explosions within a warm ionized medium are sufficient
to explain most observed characteristics of the LB. It is argued that due to 
the absence of a parent star cluster, these should be due to random explosions,
e.g. due to runaway OB stars, in the Galactic disk. As a result of the 
low energy input rate and high ambient pressure all models predict a collapse 
of interstellar bubbles once the evolution time scale is 
larger than about 6 Myr. Although an interesting idea, there is to our 
knowledge as yet no observational evidence of a bubble in contraction.

Given these widely different concepts for the origin of the LB, 
it seems appropriate to investigate the evolution of the stellar content 
of the local ISM  back in time for the last tens of million years. Thus we are
looking for a ``smoking gun'' of a stellar cluster consisting of massive
stars that may have passed through the LB; its most massive 
members should have already ended as SNe, but the later-type stars 
should still be present. Applying an initial mass function
(IMF) appropriate for OB stars we are able to infer the number
of members that may have exploded within a region that now forms the
local cavity. However, these stars may have dispersed and are not easily 
identifiable as a cluster any more. As pointed out by Eggen (1996), small 
perturbations in
the space motion of stellar clusters can lead to a significant spread of
cluster members on a relatively short time scale. In Sect.\ 2
we take a detailed look at the population of stars in the solar
neighborhood. In Sect.\ 3, we provide results of our investigations on the
youngest stars known to exist, and in Sect.\ 4 we derive the time-dependent 
energy input due to successive SN explosions and calculate the evolution 
of the LB. Finally, in Sect.\ 5, we discuss our findings
and present our conclusions on the formation of the LB.
 
\section{The local stellar population}

Investigations of the stellar content in the solar neighborhood have shown
that young stars do exist. Obviously, these stars cannot have the same origin
as the much older stellar population including our Sun. Ongoing star formation
in combination with phase-mixing and relaxation must have led to young and old
stars now co-existing in the solar neighborhood.

Detailed studies of the Hipparcos data have provided the
existence of a number of stellar moving groups. Such moving groups are star
streams identified by a similar velocity component in the direction of Galactic
rotation. A complete review of the concept of such star streams and the
Galactic structure can be found in Eggen (1996). It is suggested that some, if
not all, of these moving groups belong to so-called superclusters. According to
Eggen (1996), internal and external stresses (e.g., disk heating and Galactic
differential rotation) may cause perturbations in the space velocities of the
individual stars in a cluster, which led the original cluster to expand in
a relatively short time and form a supercluster.

Recently, Asiain et al. (1999a) studied the kinematical properties of 1924
B, A and F main sequence stars extracted from the Hipparcos catalog.
They included radial velocity measurements to derive galactic space velocity
components (U,V,W) and Str\"omgren photometry to determine stellar ages for
their sample stars. Using a method based on non-parametric density estimators
they find evidence for moving groups in the four-dimensional (U,V,W,age) space.
In addition to already known moving groups, e.g., the Sirius and the 
Pleiades moving groups, they detect finer substructures with kinematical
properties apparently consistent with those of nearby open clusters or
associations. In particular, they find that the Pleiades moving group splits
into 4 significantly different subgroups
of vastly different ages ranging from 20 to 300 Myrs. According to Asiain et
al. (1999a), the youngest (B1 in their notation) is also the
most distinct subgroup and the only one still concentrated in space. Most of
the 27 member B stars lie in the nearest direction of the Gould Belt,  
$l \approx 300\degr$. The center of mass of these stars is located 
$135 \pm 15$ pc away from the Sun, whereas 
the distance of the Gould Belt in this direction is $\sim 160$ pc. Obviously,
this Pleiades moving group is located in the foreground of the Gould Belt.
According to Asiain et al. (1999b) the average age of these stars is 
$20 \pm 10$ Myrs. Based on simulations of the evolution of a stellar complex
including the influence of the Galactic gravitational potential and the effect
of disk heating with a constant diffusion coefficient, Asiain et al. (1999b)
are also able to explain the gross properties of the kinematics of the 
Pleiades moving group substructures.

\section{The youngest stars in the solar neighborhood}
The stars associated with Pleiades subgroup B1 constitute the youngest stars
known to exist in the solar neighborhood. In order to explore whether
more massive stars existed in this association that already exploded as
SNe while moving through the LB region, we assume coeval
star formation and investigate the mass spectrum of the stars in the
Pleiades subgroup B1.

\begin{figure}
  \resizebox{\hsize}{!}{\includegraphics{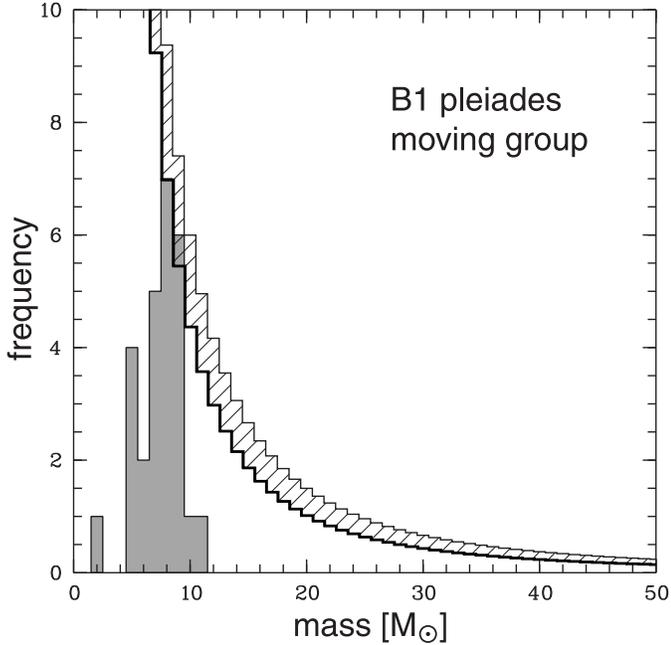}}
  \caption{Mass histogram of the stars in the Pleiades B1 moving
    group (dark shaded area). The thick solid line shows the best fit initial 
        mass function. Since the number of stars known so far to be associated
with subgroup B1 is actually a lower limit, we also plot the uncertainties 
as a hatched region (see Sect.\ 3).}
  \label{histo}
\end{figure}

Employing evolutionary tracks by Schaller et al. (1992) for the 27 stars in
subgroup B1 we derived individual stellar masses from their spectral types
and luminosities. In Fig.\ \ref{histo} we show the mass spectrum of the 
stars in this
subgroup. As can be seen, subgroup B1 is dominated by early B stars of spectral
types B3 -- B1.5, with masses in the range 6--9 M$_{\sun}$. It is
worthwhile to mention that 11 of these stars have already evolved from the
main-sequence. Note that the sample of stars investigated by Asiain et al.
(1999a) is not a complete sample. 
The apparent deficit of stars with masses less than $6 \, \msol$ in 
Fig.\ \ref{histo} can be attributed to this incompleteness. In
fact, the number of stars known so far to be associated with subgroup
B1 is a lower limit. 
In order to investigate how many stars more
massive than 9 M$_{\sun}$ may have existed in the Pleiades subgroup B1 we
assume that all stars in the mass range 8--9 M$_{\sun}$ are detected. 
Next, an initial mass function (IMF) is fit to the 
distribution of stars in the mass range 8--9 M$_{\sun}$. We adopted a
standard power law IMF with a power law index $\Gamma$ of the form
\begin{equation}
\Gamma = {d\log\xi(\log m) \over d\log m} \,;
\end{equation}
$\xi$ denotes the number of stars per unit logarithmic mass interval per unit 
area.
This translates into a number $N(m)$ of stars in the mass interval $(m, m+dm)$,
\begin{equation}
N(m) dm = N_0 \left(m\over m_0\right)^{\Gamma-1} dm \,.
\end{equation}
According to Massey et al. (1995), $\Gamma = -1.1 \pm 0.1$ for stars in 
Galactic OB associations with masses in excess of 7 M$_{\sun}$. Assuming this
value and normalizing the IMF to solar masses and the mass spectrum of
subgroup B1, $N(m_0=8\, {\rm M}_\odot) = 7$ (see Fig.\ \ref{histo}), we 
derive
\begin{equation}
N(m) = N_0 \left(m\over {\rm M}_\odot\right)^{\Gamma-1} \,, 
N_0 = 551.6^{+181}_{-103}.
\label{masdis}
\end{equation}
This IMF is shown as a thick solid line in Fig.\ \ref{histo}. Note that the 
number of 
stars known so far to be associated with subgroup B1 is actually a lower 
limit. Therefore, we also plot the uncertainties in positive direction as a
hatched region. The uncertainties take into account the uncertainties in the 
IMF slope ($\Gamma = -1.1 + 0.1 = -1.0$), which slightly flattens the IMF, and
the positive statistical number errors in the mass bins 8 and 9 $\msol$.
If we assume that such a continuous mass distribution is representative 
for the discrete number of stellar masses in a cluster, and that there 
are no stars if $N(m_{\rm max}) \leq 1$, we obtain from Eq.~(\ref{masdis}), 
$m_{\rm max} = 20.2$. Therefore, it is unlikely that stars with masses in 
excess of $\approx 20$ M$_{\sun}$ have existed in subgroup B1.
According to Fig.\ \ref{histo} mass depletion becomes apparent for stars 
with masses $m\geq m_{\rm min} = 
10\, {\rm M}_\odot$; the estimated total number of SN events, 
$N_{\rm SN}$, which will have occurred in this mass range, is then derived 
from
\begin{eqnarray}
\label{nsn1}
N_{\rm SN} &=& N_0 \times \int_{m_{\rm min}}^{m_{\rm max}} 
\left(m\over {\rm M}_\odot\right)^{\Gamma-1} dm \\
&\approx& 21.2^{+15.4}_{-4.8} \nonumber \,,
\end{eqnarray}
for $\Gamma = -1.1\pm 0.1$. Taking into account that two stars are still 
present with masses of 10--11 $\msol$, 19 SNe are estimated to have already 
occurred. Taking into account that
the sample of stars known to be associated with subgroup B1 may be incomplete,
the derived number of SN events  may be considered as a lower limit.

\begin{figure}
  \resizebox{\hsize}{!}{\includegraphics{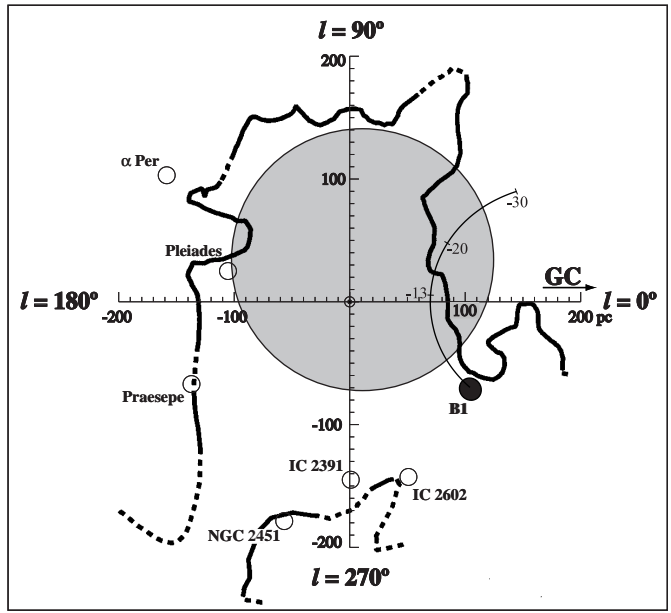}}
  \caption{Sketch of the solar neighborhood seen from above the galactic
plane. The center of mass position of Pleiades subgroup B1 is labeled with
``B1''. The solid line, ending at the actual position of B1, provides the 
trajectory of the moving group during the past 30 Myrs in the epicyclic 
approximation (see Sect.\ 3); center of mass positions 13, 20, and 30 Myrs
ago are labeled with -13, -20, and -30. Approximately 13 Myrs ago the most 
massive B1 star(s) (M $\approx 20$ M$_{\sun}$) must have exploded. 
The local cavity contours as derived from Na{\sc i} absorption line studies
by Sfeir et al. (1999) are shown as thick solid lines (dashed lines 
denote directions of uncertain local cavity borders). As can be seen, existing
B1 member stars (or at least some of them, given their spatial spread) should 
have crossed the region, which now forms the Local Bubble.
}
  \label{scetch}
\end{figure}

We next compare location and space motion of the Pleiades subgroup B1 with 
respect to the LB. Fig.\ \ref{scetch} provides a sketch of the solar 
neighborhood in a projection seen from the top of the galactic plane; the 
direction towards the galactic center is to the right ($l = 0^\circ$).
The positions of a number of nearby open stellar clusters (open circles in
Fig.\ \ref{scetch}) are shown together with the center of mass position of 
the Pleiades subgroup B1 (filled circle in Fig.\ \ref{scetch}). 
We point out that the Pleiades
subgroup B1 is localized in a small volume in space. The evolution of the space
distribution of the single member stars is plotted in detail in Asiain et al. 
(1999b, cf. Fig.\ 4).
The LB is idealized as a sphere; radius and
center have been chosen to be consistent with the EUVE measurements presented
by Bergh\"ofer et al. (1998, cf. Fig. 13). Additionally, we plot the border
of the local cavity as derived by Sfeir et al. (1999) from Na{\sc i} absorption 
line studies. 

The solid line, ending at the actual position of B1, provides the 
space motion of the moving group during the past 30 Myrs
in the epicyclic approximation; tickmarks indicate the center of mass position
30, 20, and 13 Myrs ago. The path of B1 has been computed by employing the 
equation of motion provided by Asiain et al. (1999;  Eq. (1)). The IAU 
standard for the solar motion has been adopted to transform the B1 velocity 
components in the local standard of rest (LSR) system, following the assumption
that the system of the LB is bound to the LSR. The time of -13 Myrs is 
estimated to reflect the time when the most massive star(s) in B1 
(M $\approx 20$ M$_{\sun}$) must have exploded. At this time the
center of mass of B1 is located inside the volume of the LB. 

We note in passing that there exists a controversy in the interpretation of 
Hipparcos data, resulting in different values for the solar 
motion. While there is good agreement on U$_{\sun}$, Dehnen \& Binney (1998) 
obtain a value V$_{\sun} \approx 5.25$ km\,s$^{-1}$, by arguing that 
because there is a linear relationship between ${\rm V}$ and the velocity 
dispersion $\sigma^2$ (also known as asymmetric drift), which 
is fulfilled for all stars but the youngest, that the latter ones should be 
excluded from the sample of stars. On the other hand,   
Torra et al. (2000) have analyzed the kinematics of young 
stars, using the Hipparcos data plus photometric distances, and by 
taking into account a distortion of the velocity field by the expansion of the 
Gould Belt, obtain V$_{\sun} \approx 13.4$ km\,s$^{-1}$, which is close to 
the IAU standard value of V$_{\sun} \approx 12.0$ km\,s$^{-1}$. 
A more detailed discussion of these problems is clearly beyond the scope of 
this paper.
Assuming the smaller value for V$_{\sun}$ by Dehnen \& Binney (1998) stretches 
the B1 trajectory to the upper direction in Fig.~\ref{scetch} by a factor of
$\sim$2; however, the trajectory still passes through the LB as outlined 
by the Na{\sc i} observations.

Taking into account the spread in the actual positions of individual B1 member
stars and the deviations in space velocities 
($\Delta$U, $\Delta$V) = (4.7,3.3) km\,s$^{-1}$, a significant fraction of 
former more massive members must have crossed the LB region even closer to the
center, while a few might have exploded outside. From the present positions 
of the individual stars in B1 and their
space velocities we estimate that an individual star position may deviate from
the center of mass 13 Myrs ago by up to $\sim$100 pc. Even if the 
path of B1 did not run through the very center of the LB it
should be emphasized (and will also be discussed in the next section) that
the shape or boundary of the bubble are mainly determined by the density and 
pressure of the {\em ambient 
medium} rather than by the location of individual SNe with respect to the 
center of explosion. The reason is that a shock wave is weak and propagates 
fast in a rarefied hot medium, but moves rather slowly when it encounters 
high density material, which it has to compress and set into motion. 

\section{Energy budget and evolution of the Local Bubble}
While the details of the dynamics and evolution of the LB 
have to be deferred to a future paper, which concentrates  
on numerical simulations, we will discuss in the following the 
time-dependent energy input rate and its consequences for 
the evolution of the LB. 

Obviously,
the contribution by stellar winds is only significant at early times and thus
is relevant only if massive O stars exist in the cluster, which is not the 
case here. 
We will therefore only consider the input rate by SN events, 
which is by far the dominant at later times (Mac Low \& McCray 1988). 
The main sequence life times, $\tau_{\rm ms}$, of stars within 
the mass range $7\msol \leq m \leq 30 \, \msol$ can empirically be 
approximated by $\tau_{\rm ms} = 3\,10^7 \, (m/[10 \msol])^{-\alpha}$ yr 
(Stothers 1972), with $\alpha = 1.6$. 
Since this defines $m$ as a function of time $\tau$, implicitly 
assuming that the energy input can be described as a continuous process, 
we obtain
\begin{equation}
{m\over {\rm M}_\odot} = \left(\tau \over C\right)^{-1/\alpha} \,, 
\label{tms}
\end{equation}
with $C = 3.762 \times 10^{16}$ s.
It should be borne in mind that describing discrete events (i.e.\ SN 
explosions) by a continuous mass distribution function in time may disregard 
considerable fluctuations in the time-dependent quantities calculated below.
However, the results obtained should be accurate enough in a time averaged 
sense. 
Let $L_{\rm SB}$ be the energy input rate per unit time due to a 
number of successive SN explosions but with a constant energy input 
of $E_{\rm SN} = 10^{51}$ erg each, then the cumulative number of SNe 
down to mass $m$ reads
\begin{eqnarray}
N_{\rm SN}(m) = 501.45 \, \left(m\over {\rm M}_\odot\right)^\Gamma - B \,,
\end{eqnarray}
with $B= 18.58$, and 
\begin{eqnarray}
L_{\rm SB} &=& {d \over dt} (N_{\rm SN} E_{\rm SN})= E_{\rm SN} 
{d N_{\rm SN} \over dt} \nonumber\\
&=& E_{\rm SN} {d N_{\rm SN} \over dm} {dm \over d\tau} {d\tau \over dt} 
\,; 
\end{eqnarray}
using Eqs.~(\ref{nsn1}) and (\ref{tms}), 
\begin{equation}
L_{\rm SB} = 344.75 \, C^{\Gamma/\alpha} t^{-(\Gamma/\alpha + 1)} \,;
\label{eninp1}
\end{equation}
with the above values for $\Gamma$ and $\alpha$, we obtain the handy 
formula
\begin{equation}
L_{\rm SB} = L_0 \, t_7^{\delta} \,,
\label{eninp2}
\end{equation}
where $L_0 = 4.085 \times 10^{37} \,{\rm erg/s}$,  
$\delta = -(\Gamma/\alpha + 1) = -0.3125$ and $t_7 = t/10^7$ yr.
Here and in the following, the variable $t$ has been used as the dynamical 
time for energy input into the LB and its evolution. 
Assuming coeval star formation we estimate from the lack of stars with 
$m \geq 10 - 11 \, {\rm M}_\odot$ the age of the star cluster to be
$\tau_{cl} \leq 2.5 \times 10^7$ yr; good agreement with the evolution 
tracks by Schaller (1992) is obtained for $\tau_{cl} = 2.3 \times 10^7$ 
(s.~Fig.~\ref{scetch}).
The present dynamical age for the bubble evolution is therefore given by  
$t_0 = \tau_{cl} - \tau_0$ $= 1.3 \times 10^7$ yr, where $\tau_0 = 
\tau_{ms}(m_{max}) \approx 9.9$ Myr defines point zero as the main sequence 
life-time for the most massive star ($m_{max} \sim 20 \, {\rm M}_\odot$). 
The dynamical time is then given by $t = \tau - \tau_0$, and hence 
$d\tau = dt$.

The assumption of a continuous energy input into the ISM at the 
{\em center} of the first explosion is much better than thought of at first 
glance. The reason is that every off-center blast wave in a hot rarefied 
medium with a high speed of sound rapidly degenerates into a Mach 1 shock 
with a low factor of compressibility of the downstream medium. Therefore, 
hardly any material is swept up and the bubble remains roughly uniform 
with negligible radiative cooling due to its high initial temperature. The net 
effect is simply pressurization of the bubble. The time for the interior 
weak shock to reach the bubble boundary is of the order of a sound crossing 
time and thus during most of the evolution much smaller than the dynamical 
time scale; hence the eccentricity of subsequent explosions has only a minor 
influence on the evolution of the bubble.  

Similarity solutions of superbubbles (e.g.\ McCray \& Kafatos 1987) and 
their classical analogue, stellar wind bubbles (e.g.\ 
Pikelner 1968, Dyson \& deVries 1972, Castor et al. 1975, 
Weaver et al. 1977), can be calculated, provided that the dependent 
variables and boundary conditions do not explicitly depend on time 
or length scales that may enter the problem. This is still true for 
the case of a time-dependent energy input rate, $L_{\rm SB} = L_0 t^\delta$ 
and a density gradient of the ambient medium, 
$\rho(r) = \rho_0 r^{-\beta}$. 

Similarity conditions hold during the adiabatic expansion phase as long 
as the bubble pressure is much larger than the ambient medium pressure, 
i.e.\ $P_b \gg P_0$. The mass of the bubble is dominated by the mass 
swept into a shell, 
\begin{equation}
M_{sh}(r) = \int^r \rho(r^\prime) d^3 r^\prime \,,
\label{mascon1}
\end{equation}
and the energy input is 
shared between kinetic and thermal energy, 
\begin{equation}
E_{th}(r) = 1/(\gamma -1) \int^r p(r^\prime)  d^3 r^\prime \,.
\end{equation}
Using $\gamma = 5/3$ for the ratio of specific heats, observing that 
$P_b$ remains uniform, and applying spherical symmetry, conservation 
of momentum 
\begin{equation}
{d\over dt} (M_{sh} \dot R_b) = 4 \pi R_b^2 P_b \,,
\end{equation}
and energy
\begin{equation}
{d E_{th} \over dt} = L_{\rm SB}(t) - 4 \pi R_b^2 \dot R_b P_b \,,
\end{equation}
yields the solution 
\begin{eqnarray}
R_b &=& A t^\mu \, \\
A &=& \left\{{(5-\beta)^3 (3-\beta) \over (7 \delta - 
\beta - \delta \beta + 11) (4 \delta + 7 - 
\delta \beta - 2 \beta)}\right\}^{1/(5-\beta)} \times \nonumber\\ 
&& \times \left\{{L_0 \over 2\pi (\delta + 3) \rho_0} \right\}^{1/(5-\beta)} \, \\
\mu &=& {\delta + 3 \over 5 - \beta} \,;
\label{simsol1}
\end{eqnarray}
since the swept-up shell is usually thin, the bubble and shell radius 
can be treated as equal during the energy driven phase and are denoted by 
$R_b$.
The similarity variable in the case considered here is given by
\begin{equation}
\mu = {2-\Gamma/\alpha \over 5-\beta} \,.
\end{equation}
To keep matters simple, the ambient density is assumed to be constant 
($\beta=0$), although it is possible that some of the explosions 
occur inside a molecular cloud (s. Breitschwerdt \& Schmutzler 1994). 
With the previously used values, the radius of the bubble evolves as
\begin{equation}
R_b = 251 \left(2 \times 10^{-24} {\rm g}/{\rm cm}^3 \over \rho_0\right)^{1/5} 
      t_7^{0.5375} \, {\rm pc} \,,
\label{bubrad1}
\end{equation}
and the expansion velocity as 
\begin{equation}
\dot R_b = 13.22 \left(2 \times 10^{-24} {\rm g}/{\rm cm}^3 
           \over \rho_0\right)^{1/5} t_7^{-0.4625}\, {\rm km/s} \,.
\label{bubvel1}
\end{equation}
As a consequence of a decreasing energy input rate the expansion law exponent 
in Eq.~(\ref{bubrad1}) is between a Sedov ($\mu=0.4$) and a stellar wind 
($\mu=0.6$) type solution. 
According to Eqs.~(\ref{bubrad1}) and (\ref{bubvel1})
the radius of the LB will be 289 pc and 158 pc and its velocity is 11.7 
km/s and 6.4 km/s, if the 
ambient density is $\rho_0 = 2 \times 10^{-24} \, {\rm g}/{\rm cm}^3$ and  
$\rho_0 = 4 \times 10^{-23} \, {\rm g}/{\rm cm}^3$, or 
$n_0 = 1 \, {\rm cm}^{-3}$ and $n_0 = 20 \, {\rm cm}^{-3}$, respectively. 
In the latter case the value of the ambient density 
would correspond roughly to that of the cold neutral medium. 
It is interesting to note that the radius, the dynamical time or age  
of the LB, and the rate of SNe ($f_{SN} \sim 1/(6.5 \times 10^5 \, 
{\rm yr})$) are very similar to that of the neighboring Loop I superbubble; 
Egger (1998) estimated its radius to be $R_{LI} = 158$ pc, its dynamical 
age to be $t_{dyn} \sim 10^7$ yr and the SN rate to be  
$f_{SN} \sim 1/(6 \times 10^5 \, {\rm yr})$. Loop~I is an active superbubble 
and has been powered by about 40 SNe in the Sco-Cen OB association, which 
is about twice as many as in the case of the LB; in this 
respect, its radius is also fairly small. 
 
According to ROSAT PSPC broad band spectral fitting the Loop I bubble 
temperature 
is about $2.5 \times 10^6$ K (Egger 1998). This gives support to the 
idea that we live in an environment of ongoing star formation, although 
the LB itself is probably extinct.

There are several reasons why we may have overestimated the size of the LB 
in the similarity solutions above. Firstly, the mass inside the bubble 
is significantly higher than the pure ejecta mass, as can be inferred 
from the ROSAT X-ray emission measures; when assuming bubble parameters 
of $R_b = 100$ pc and $n_b = 5 \times 10^{-3} \, {\rm cm}^{-3}$ (e.g.\ 
Snowden et al. 1990) a mass of at least 600 ${\rm M}_\odot$ is derived whereas
non-equilibrium ionization plasma models (Breitschwerdt \& Schmutzler 1994)
result in a more than a factor of five higher mass.

The contribution of ejecta is of the order of 100 ${\rm M}_\odot$, and the 
bulk of the bubble mass is due to hydrodynamic mixing of shell material, 
heat conduction between shell and bubble and evaporation of entrained 
clouds; therefore the flow is mass-loaded. 
The net effect is to reduce the amount of specific energy per unit 
mass, because the material mixed in is essentially cold, and thereby also 
increases the rate of radiative cooling. Secondly, the 
stellar association has probably been surrounded by a molecular cloud 
with a density in excess of $n_0 = 100 \, {\rm cm}^{-3}$ with subsequent 
break-out of the bubble and dispersal of the parent cloud 
(Breitschwerdt et al. 1996). Thirdly, the number of SN explosions could 
be less; here we have assumed that all 20 SNe have 
occurred inside the LB. This need not be the case as is also suggested 
by Fig.\ \ref{scetch}. ROSAT PSPC observations have revealed an annular 
shadow centered toward the direction ($l_{\rm II} = 335^\circ$, 
$b_{\rm II} = 0^\circ$), which has been interpreted as an interaction 
between the LB and the neighboring Loop I superbubble (Egger \& Aschenbach 
1995). The trajectory of the cluster B1 may have partly crossed the Loop I 
region. Alternatively and more likely, part of the thermal energy might 
have been liberated into the Galactic halo, since there is some evidence 
(see Sect.~\ref{disc}) that the LB is open toward the North Galactic 
Pole. It should also be mentioned that due to small number statistics  
the true number of SNe can vary by a factor of 2. 
Finally, although there is no stringent evidence, it would be very 
unusual, if the LB would not be bounded by a magnetic field, whose 
tension and pressure forces would decrease the size of the LB. 

Given these uncertainties and the fact that the simple analytic model 
discussed above can be considered as an upper limit, the agreement with 
observations is quite good. The bubble radius and shell velocity are 
rather insensitive to the energy input rate and the ambient density 
(due to the power of $1/5$) and therefore not well constrained, but 
depend more sensitively on the expansion time scale. Thus we can assert 
with some confidence that the age of the LB should be between 
$1 - 2 \times 10^7$ yr.   

The radiative cooling time scale of the LB can be estimated using 
Kahn's (1976) cooling law (in the temperature range 
$10^5 \leq T \leq 10^7$ K), 
$t_c = \kappa^{3/2}/q$, with $\kappa = 
P/\rho^{5/3}$ being the adiabatic parameter and 
$q = 4 \times 10^{32} \, {\rm cm}^6/({\rm s}^4 {\rm g})$. Using $n_b = 
5 \times 10^{-3}$ and $T_b = 10^6$ K, which have been derived from 
X-ray broad band spectral fitting of a plasma in collisional 
ionization equilibrium (e.g.\ Snowden et al. 1990), we obtain 
$t_c \approx 1.3 \times 10^7$ yr, comparable to the dynamical time scale. 
If the X-ray emission is mainly due to the delayed recombination of highly 
ionized species of a plasma not in in ionization equilibrium, then the cooling 
time can be as short as $6 \times 10^5$ yr (Breitschwerdt et al. 1996). 
In either case, the energy input of previous SNe has largely been used 
up for expansion and radiative cooling of the bubble.
We thus conclude that the emergent X-ray emission must be largely due 
to the energy input of the {\em last supernova}. 

We can express the number of stars (and potential SNe) in the mass interval 
$(m, m+dm)$, $N(m)$, by a number of stars ending their lives in the time 
interval $(\tau, \tau+d\tau)$, $N(\tau)$, using Eq.\ (\ref{masdis}) and 
$\tau = t + \tau_0$ ($\tau_0 \approx 10^7$ yr), which results in
\begin{eqnarray}
N(\tau) &=& 9.697\, 10^{-20} \times \left(\tau/[{\rm sec}]
\right)^{(1-\Gamma)/\alpha} \,, \nonumber\\
&=& 1.03 \times \left(\tau \over 10^7 \, {\rm yr}\right)^{1.3125} \\
&\approx& 1.03 \times (1+t_7)^{1.3125} \,. \nonumber
\end{eqnarray}
The average energy input rate by recent SNe, $\dot E_{\rm SN}$, (not to be 
confounded with the cumulative rate of Eq.~(\ref{nsn1})) is then given 
by
\begin{eqnarray}
\dot E_{\rm SN} &=& E_{\rm SN} {dN(\tau)\over d\tau}{d\tau \over dt} \nonumber\\
&\approx& 1.27 \, 10^{32} \times 
\left(\tau/[{\rm sec}]\right)^{(1-\alpha-\Gamma)/\alpha} \\
&\approx& 4.32 \times 10^{36} (1+t_7)^{0.3125} \nonumber \, {\rm erg/s} \,.
\end{eqnarray}
As a result of the increasing number of exploding stars with time 
(compensating their increasing main sequence life time) and the
assumption that the individual contribution is constant, i.e.\
$E_{\rm SN} = 10^{51}$ erg, the energy input rate (using $\Gamma = -1.1$) 
is a moderately increasing function with time.

Analysis of ferromanganese crust samples in deep ocean layers have 
shown an increased flux rate of ${}^{60}{\rm Fe}$, which is indicative of 
at least one SN which exploded inside the LB 
at about $5$ Myr ago (Knie et al. 1999). Thus, using $t_7 = 0.8$, we obtain 
for the associated energy input rate, 
$\dot E_{\rm SN} = 5.18 \times 10^{36}$ erg/s.
This is a moderate value, but still comfortably larger by a factor of 5  
than the total present day soft X-ray photon energy output rate of the LB.  

\section{Discussion and Conclusions}
\label{disc}
The fact that the Gould Belt stars show receding velocities, and
that the stellar associations are also closely related to
molecular clouds and local H{\sc i} (see P\"oppel 1997) argues
in favor of a common
history of gas and stars in the local environment.
The nature of the Gould Belt is yet unclear;
current ideas favor either an explosive origin (Blaauw 1965, Olano 1982)
or a collision between a high velocity cloud and the Galactic disk
(e.g.\ Tenorio-Tagle et al. 1986; Comer\'on \& Torra 1992).
It is noteworthy that the LB extends
perpendicular to the Gould Belt rather than to the Galactic disk (Sfeir
et al. 1999), hinting at an origin of the LB connected to the 
Gould Belt stars and gas.

Regardless
whether the expansion was caused by multiple SN events or by the
oblique impact of a high velocity cloud,
the implications are similar: the shock fronts that
have propagated into the ISM, presumably at some angle with respect
to the disk, will overrun ambient clouds. If the clouds
have low masses, they could be disrupted hydrodynamically and/or
evaporated, thereby mass loading and decelerating the flow.
As a matter of fact, the spectral properties of the soft X-ray emission 
combined with the H{\sc i} distribution in the local cavity show that the hot 
LB plasma must consist of a total mass of the order of 
$1000 \, {\rm M}_\odot$. This cannot
be provided solely by ejecta material, but must be due to hydrodynamic
mixing and/or evaporation of ambient gas clouds.
In the case of more massive clouds, the
shock propagating into the gas will compress it
without destroying the cloud, so that
a growing number of Jeans unstable clumps will eventually collapse
into protostellar cores. A considerable number of nearby stars with
ages substantially lower than the Gould Belt's kinematical age and the 
existence of active superbubbles (e.g. Loop I with an estimated dynamical age 
of $10^7$ years, see Egger 1998) seem to support this hypothesis. 
Note that the age of the Gould Belt is rather uncertain, Comer\'on (1999) 
provides an age estimate of $34 \pm 3$ Myrs whereas Torra et al. (2000) derive
an age in the range of 30--60\,Myrs.

In the above picture, the origin of the LB is connected to the evolution of 
the Gould Belt itself as the disturbed medium that formed the LB. 
We favor a scenario in which the LB was created by about $10 - 20$ SNe 
about $1 - 2 \times 10^7$ yr ago, with a preferred value of 
$1.3 \times 10^7$ yr; the star cluster itself is about $2.3 \times10^7$ yr.    
It is conceivable that the stars have also been embedded in a 
molecular cloud for some time (Breitschwerdt et al. 1996).  
The present day total LB X-ray luminosity
is fairly moderate, $L_X \sim 10^{36} \, {\rm erg/s}$ at most
(Cox \& Reynolds 1987), and can easily be supplied by the last SN
reheating the cavity about a million years ago. This could also explain 
why only part of the local cavity is illuminated now by X-rays. Note 
that the sound crossing time out to a distance of 200 pc (in direction of 
$\beta$CMa) in a $10^6$ K plasma is roughly 2 Myrs. Moreover, the 
recombination time scale of major coolants, such as O{\sc vii} and 
C{\sc vi} can be estimated from analytical approximations to the rate 
coefficients (Shull \& van Steenberg 1982) to be $4.7$ and $12.2$ Myrs, 
respectively, for commonly used LB fit parameters of $n_b = 5 \times 10^{-3} 
\, {\rm cm}^{-3}$ and $T_b = 10^6$ K (Breitschwerdt \& Freyberg 2002). 
Therefore the emergent X-ray spectrum should be a fairly complex superposition 
of freshly ionized and recombining plasma (from previous explosions) and 
definitely be out ionization equilibrium.

We have shown that the subgroup B1 of the Pleiades moving group
has actually passed through the LB. If the IMF of this
cluster of young stars is similar to the OB associations we observe in
our local neighborhood, then obviously some of
these stars must have exploded as SNe on their way. Since the kinematic age
of B1 can be fairly well determined (assuming they are not bound
gravitationally) to be $\tau _{B1} \sim 20$ Myrs,
it can certainly not be responsible for the origin of the Gould Belt
itself. Note that all age determinations for the Gould Belt point to ages
older than 30 Myrs. However, the SN events in B1 should have left their 
imprint on the LB. One important question is when the
last one injected mass and energy. According to Eq.\ (\ref{masdis}) 
and Fig.~\ref{histo} about 4--5
stars of $10 - 11\, \msol$ should have been members of B1. 
Since two of them are 
still with us and the average life time of such stars is about 25 Myrs, 
while stars with $9 - 10\, \msol$ live 30 Myrs
the last SN is expected to have exploded about $1 - 2$ Myrs ago, or 3 SNe 
occurred within the last 5 Myrs. This is roughly consistent with the 
measurements of radioactive ${}^{60}{\rm Fe}$ (Knie et al. 1999), where 
the high flux rate is interpreted by one SN, but within a radius of 30 pc 
from the Sun and a progenitor mass of about $15 \, \msol$.  

The existence of a ``Local Chimney'' as claimed by Sfeir et al. (1999) and
Welsh et al. (1999) may support the hypothesis that the LB is created and 
shaped by as many as 20 SNe, because it allows a significant fraction of the 
large superbubble thermal energy to be vented into the Galactic halo. 
Consequently, after the last explosion, the LB is left in a state of long-term 
recombination.
According to non-equilibrium models (Breitschwerdt \& Schmutzler 1994; 
Breitschwerdt 2002) the observed modest amount of soft X-ray emission can still
be generated, even if the last SN occurred more than a million years 
ago. 

B1 still provides another 10 SN candidate members. Since it is about to 
leave the volume of the local cavity, the LB will remain in a state of
recombination, which is what we probably observe at present. Reheating will 
become progressively less important and the future of 
the LB should be fairly quiescent.

\begin{acknowledgements}
TWB acknowledges support from the
Alexander-von-Humboldt-Stiftung (AvH) by a Feodor-Lynen Fellowship.
The research of DB has been funded by the Deutsche 
Forschungsgemeinschaft (DFG) with a
{\em Heisenberg Fellowship} and by the Max-Planck-Gesellschaft (MPG). 
DB thanks G.\ Hasinger, J.\ Tr\"umper and the Max-Planck-Institut f\"ur 
extraterrestrische Physik as well as the Department of Astrophysical Sciences
at Princeton University, where part of this work was done, for their
hospitality. 

We are indebted to F.\ Figueras and R.\ Asiain of the
Departament d'Astronomia i Meteorologia, Universitat de Barcelona, who
kindly provided us with the listings of stars in the Pleiades moving groups,
and to D. Fernandez for helping us in calculating the trajectories of B1.
We kindly acknowledge valuable discussions with F. Comer\'on (ESO), B. Fuchs
(ARI Heidelberg) and C. de Vegt (Hamburger Sternwarte).
\end{acknowledgements}

\end{document}